\documentclass[a4paper, fleqn, usenatbib, useAMS]{mnras}

\usepackage{graphicx}	% Including figure files
\usepackage{amsmath}	% Advanced maths commands
\usepackage{amssymb}	% Extra maths symbols
\usepackage{multicol}   % Multi-column entries in tables
\usepackage{bm}		% Bold maths symbols, including upright Greek
\usepackage{pdflscape}	% Landscape pages
\usepackage{subfig}	% 
\usepackage[normalem]{ulem}
\usepackage{xspace}

\newcommand{\Msun}{\ensuremath{M_{\odot}}}%
\newcommand{\Lsun}{\ensuremath{L_{\odot}}}%
\newcommand{\mags}{mag\,arcsec\ensuremath{^{-2}}\xspace}%
\newcommand{\re}{\ensuremath{r_\mathrm{e}}\xspace}%
\newcommand{\rO}{\ensuremath{r_\mathrm{0}}\xspace}%
\newcommand{\mue}{\ensuremath{\mu_\mathrm{e}}\xspace}%
\newcommand{\muO}{\ensuremath{\mu_\mathrm{0}}\xspace}%
\newcommand{\Ie}{\ensuremath{I_\mathrm{e}}\xspace}%
\newcommand{\IO}{\ensuremath{I_\mathrm{0}}\xspace}%
\newcommand{\me}{$\langle \epsilon \rangle$\xspace}%
\newcommand{\mat}{$\langle a_{3}/a \rangle$\xspace}%
\newcommand{\mac}{$\langle a_{4}/a \rangle$\xspace}%
\newcommand\nodata{ ~$\cdots$~ }%

\usepackage[T1]{fontenc}
\usepackage{ae,aecompl}

\usepackage{newtxtext, newtxmath}

\graphicspath{{figures/}}
\title[Antlia cluster: Catalogue and isophotal analysis]{%
  Early-type galaxies in the Antlia Cluster:\\
  Catalogue and isophotal analysis}%
\author[J. P. Calder\'on et\,al.]{%
  Juan P. Calder\'on$^{1,2,3}$\thanks{E-mail:\,jpcalderon@fcaglp.unlp.edu.ar}, %
  Lilia P. Bassino$^{1,2,3}$, Sergio A. Cellone$^{1,3,4}$ and %
  Mat\'ias G\'omez$^{5}$\\
  $^1$Consejo Nacional de Investigaciones Cient\'ificas y T\'ecnicas, %
  Rivadavia 1917, Buenos Aires, Argentina\\
  $^2$Instituto de Astrof\'isica de La Plata (CCT La Plata - CONICET - UNLP), %
  La Plata, Argentina\\
  $^3$Facultad de Ciencias Astron\'omicas y Geof\'isicas, Universidad Nacional %
  de La Plata, Paseo del Bosque, B1900FWA La Plata, Argentina\\
  $^4$Complejo Astron\'omico El Leoncito (CONICET - UNLP - UNC - UNSJ), %
  San Juan, Argentina\\
  $^5$Departamento de Ciencias F\'isicas, Facultad de Ciencias Exactas, %
  Universidad Andres Bello, Santiago, Chile\\
}

\begin{document}
\date{...}
\pagerange{\pageref{firstpage}--\pageref{lastpage}} \pubyear{2018}

\maketitle
\label{firstpage}
\begin{abstract}
  We present a statistical isophotal analysis of 138 early-type
  galaxies in the Antlia cluster, located at a distance of
  $\sim$\,35\,Mpc. The observational material consists of CCD images
  of four 36\,arcmin\,$\times$\,36\,arcmin fields obtained with the
  MOSAIC\,II camera at the Blanco 4-m telescope at CTIO.  Our present
  work supersedes previous Antlia studies in the sense that the
  covered area is four times larger, the limiting magnitude is $M_{B}
  \sim -9.6$\,mag, and the surface photometry parameters of each
  galaxy are derived from S\'ersic model fits extrapolated to
  infinity. In a companion previous study we focused on the scaling
  relations obtained by means of surface photometry, and now we
  present the data, on which the previous paper is based, the
  parameters of the isophotal fits as well as an isophotal analysis.

  For each galaxy, we derive isophotal shape parameters along the
  semi-major axis and search for correlations within different radial
  bins. Through extensive statistical tests, we also analyse the
  behaviour of these values against photometric and global parameters
  of the galaxies themselves.
  
  While some galaxies do display radial gradients in their ellipticity
  ($\epsilon$) and/or their Fourier coefficients, differences in mean
  values between adjacent regions are not statistically
  significant. Regarding Fourier coefficients, dwarf galaxies usually
  display gradients between all adjacent regions, while non-dwarfs
  tend to show this behaviour just between the two outermost regions.
  Globally, there is no obvious correlation between Fourier
  coefficients and luminosity for the whole magnitude range ($-12
  \gtrsim M_V \gtrsim -22$); however, dwarfs display much higher
  dispersions at all radii.
\end{abstract}


\section{Introduction}\label{sec:introduction}
Since the early work of \cite{1968adga.book.....S}, the study of the
surface brightness profiles of elliptical galaxies (E) has reached a
state in which peculiarities are more the rule than the
exception. Even long-considered `canonical' examples of purely
elliptical shape like NGC\,3379 (see
e.g. \citealt{1994AJ....108..111S}) are nowadays understood as prime
focus for isophote twisting, large shells and arcs and complex
structure extending many effective radii; these evidences cast serious
doubts on the existence of alleged pure E as a class.

Even for E galaxies with symmetrical isophotes, there is usually extra
light that distorts the profile \citep[e.g.][]{1983ApJ...274..534M,
  1988ApJ...328...88S, 1990dig..book..270S, 1992ARA&A..30..705B}.
Thus, in many cases, the isophotes of these galaxies deviate
systematically from pure ellipses. Depending on the shape of those
deviations, they are referred to as `discy' or `boxy' isophotes. Discy
isophotes are the consequence of light excesses along the main axes
(major and minor) with respect to a perfectly elliptical, while boxy
isophotes are the consequence of deformations along directions at
$45^{\circ}$ from the main axes. In fact, galaxies within these two
types of isophote classifications present quite different
characteristics, defining two `families'. Boxy early-type galaxies
(ETGs) are usually luminous and massive, have significant radio and
X-ray emission, have `core' nuclear profiles and slow rotation; discy
ETGs, in turn, tend to be fainter, have significant rotation, and no (or
faint) X-ray or radio activity \citep{1994AJ....108.1598F,
  1994AJ....108.1579V, 2001AJ....121.2431R, 2005AJ....129.2138L}.

The analysis of possible correlations between isophotal shapes and
other parameters that characterise the isophotes, or the properties of
the galaxies themselves, has been the subject of many studies.
\cite{1989A&A...217...35B} and \cite{1989A&A...215..266N}, two seminal
papers on the subject, performed detailed studies of the shapes of
isophotes of massive E galaxies, and concluded that there is no strong
correlation with any photometric parameter like effective radius or
surface brightness. More recently, \cite{2013MNRAS.433.2812K} analysed
the nuclear slope of 135 ETGs and found no evidence of bimodality
regarding boxy or discy isophotes. Using the integral-field
spectroscopy obtained by the ATLAS$^{3D}$ survey,
\cite{2011MNRAS.414..888E} also pointed out that the $a_{4}/a$
parameter, i.e. the Fourier coefficient that defines
`disciness/boxiness', is not directly related with any kinematic
properties in their sample of 260 ETGs. However, galaxies surrounded
by X-ray haloes have generally irregular or boxy-type isophotes.
\cite{1989A&A...217...35B} found that boxy galaxies have higher
mass-luminosity ratios ($M/L \sim 11.5 \pm 0.9 \Msun/\Lsun$) than
discy-type galaxies ($M/L \sim 6.4 \pm 0.6 \Msun/\Lsun$). Regarding
galaxy luminosity, the fainter galaxies tend to be discy, while those
with higher luminosities tend to be boxy. These observed correlations
mark the cause of the dichotomy between the isophotes shapes and its
relation with galaxy formation history
\citep{1997ApJ...478L..17B}. Also, there is growing evidence of a
correlation between the age and the shape of galaxies, in the sense
that core Es have older stellar populations than power-law ones
\citep{2001MNRAS.326.1141R}.  In addition, \cite{2014RAA....14..144H}
investigated the relationships among isophotal shapes, galaxy
brightness profile and kinematic properties of a sample of ETGs from
DSS Data Release 8 with kinematic properties available from the
ATLAS$^{3D}$ survey. They found no clear relation between the S\'ersic
index and isophotal shape. Instead, they found correlations between
the Fourier coefficient $a_{4}/a$, ellipticity, and specific angular
momentum $\lambda{r_\mathrm{e}/2}$ for power-law galaxies, while no
relation was found for core ETGs.

From the theoretical side, there have been many attempts to understand
the origin of discy and boxy Es. \citet[][and references
therein]{2006ApJ...636L..81N} used semi-analytical simulations to
conclude that discy Es are mainly produced by non-equal mergers of two
disc galaxies, while equal-mass mergers tend to produce boxy Es.  In
addition, \cite{2005MNRAS.359.1379K} concluded that the isophotal
shapes of merger remnants also depend on the morphology of their
progenitors and the subsequent gas infall.

Our present study focuses on the Antlia cluster, which is recognised
as the third nearest rich galaxy cluster, after Fornax and Virgo. Its
galaxy population ranges in luminosity between -12 and -22\,mag in the
$T_{1}$-band, while no study of the relationship between their
isophotes and global parameters has still been done. The first study
of its galaxy content was performed by \cite{1990AJ....100....1F}, who
constructed the photographic catalogue FS90. On the basis of CCD
images, a deeper analysis of the ETGs located at the central zone of
Antlia was performed \citep{2008MNRAS.386.2311S, 2008MNRAS.391..685S,
  2012MNRAS.419.2472S}.  In the present work, we extend the studied
region approximately four times, determining total (not isophotal)
magnitudes and colours. Structural parameters have also been obtained
by means of S\'ersic model fits. Half of the studied galaxies are
included in the FS90 catalogue and the rest, mostly in the fainter
regime, are new ones. The total sample amounts to 138 ETGs, 59 of them
being spectroscopically confirmed Antlia members. These data have
already been used in a previous companion paper
\citep{2015MNRAS.451..791C}, to study the Antlia galaxies scaling
relations.

This paper presents the catalogue of structural parameters of ETGs in
the Antlia cluster and, on the basis of these data, an isophotal
analysis of the galaxy sample is made. The paper is organized in the
following way: in Section~\ref{sec:data} we describe the imaging data
reduction, while the galaxy sample selection is briefly presented in
Section~\ref{sec:the_galaxy_sample}.
Section~\ref{sec:isophotal_analysis} presents the computation of the
geometrical parameters, while in
Section~\ref{sec:surface_brightness_profiles} we describe the surface
photometry method used to obtain each galaxy profile. Our results are
presented in Section~\ref{sec:results}, and we discuss them in
Section~\ref{sec:discussion}. The main conclusions are contained in
Section~\ref{sec:summary}. The full catalogue is available in
electronic format.

\section{Data}\label{sec:data}
\begin{figure}
 \includegraphics[width=0.95\columnwidth]{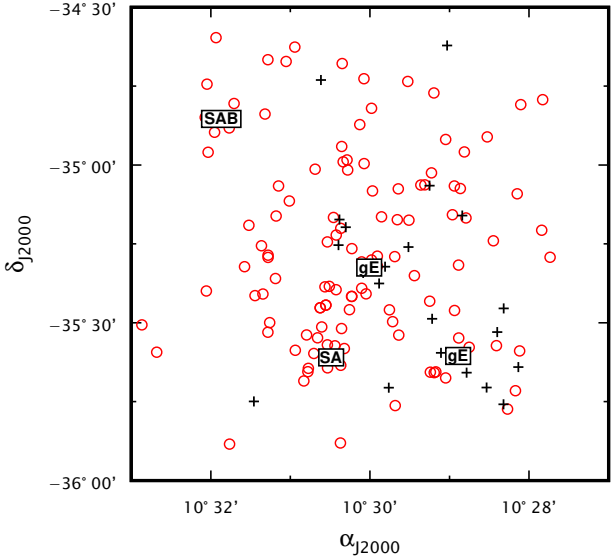}
 \caption{Projected distribution of all galaxies in the sample. 
   The faintest galaxies in the sample
   (dE and dSph) are indicated with red circles, and the brightest
   ones with black crosses. Four of the most luminous galaxies in the
   cluster are labeled with their morphologies: NGC\,3268 (`gE', at
   the centre of the figure), NGC\,3258 (`gE', at south-west),
   NGC\,3281 (`SAB') and NGC\,3273 (`SA').
   North is up and East to the left. } \label{fig:radec}
\end{figure}

The photometric data used in this paper are CCD images obtained with
the MOSAIC\,II camera, mounted on the Victor Blanco 4-m telescope at
the Cerro Tololo Interamerican Observatory (CTIO, Chile). We used the
Kron-Cousins $R$ and Washington $C$ filters
\citep{1976AJ.....81..228C}.  The $R$ filter was chosen instead of the
original Washington $T_{1}$ because of its better efficiency
\citep{1996AJ....111..480G}, while just a small change of zero-point
($R - T_{1}= 0.02$) is needed to transform between them
\citep{2003A&A...408..929D}. Each image covers 36\,arcmin $\times$
36\,arcmin, that corresponds to about $370 \times 370$\,kpc$^2$
according to the adopted Antlia distance \citep[$d=35$\,Mpc; $m-M=
  32.73$]{2003A&A...408..929D}. The MOSAIC\,II camera had a resolution
of 0.27\,arcsec/pixel and was constituted by 8 CCDs. In order to erase
the gaps between the CCDs, it is necessary to take a series of
slightly shifted exposures (dithering) and then combine
them. Figure~\ref{fig:radec} shows the projected spatial distribution
of the four MOSAIC fields used in this work, in the $R$ band. Red
circles represent the faintest galaxies in the sample (dE and dSph),
while black crosses indicate the brightest ones. We also added the
location of the more luminous galaxies in the sample: NGC\,3258,
NGC\,3268, NGC\,3281 and NGC\,3273.  We have already described the
images in \cite{2015MNRAS.451..791C}, as well as the calibration to
the standard system and the resulting signal-to-noise ratio (see
Section~\ref{sec:signal-to-noise}) of the brightness profiles, which
is extremely relevant to low surface-brightness galaxies. As a
consequence, we briefly highlight here the most important steps of
images´ reduction, as they may be of interest.

The MOSAIC\,II images reduction was made using the \textsc{mscred}
package within IRAF, which has been written specially for data of
similar characteristics \citep{1997ASPC..125..455V}. The first step
consisted in running the task \textsc{ccdproc} on all the images, in
order to perform the basic calibration (overscan subtraction,
trimming, bad pixel replacement, zero level subtraction, and
flat-fielding). As we are using images with a large field of view
(FOV), it is necessary to have an accurate celestial coordinate
system. Then, to correct the astrometric solution we ran the
\textsc{msccmatch} task, that uses a list of reference celestial
coordinates of stars located in the field, to match against the same
objects on the MOSAIC images. A polinomial relation
between the observed positions and the reference coordinates is
obtained. This relation may include a zero point shift, a scale
change, and axis rotation for both coordinate axes. Next, the fit was
applied to the multi-extension images and, using \textsc{mscimage}, it
was possible to get an output image in the correct WCS (World
Coordinate System). If any residual large-scale gradients were present
in the sky background of individual exposures, they were removed using
\textsc{mscskysub}.  In the following step, we used \textsc{mscimatch}
to match the intensity scales on the different images to be finally
combined into the stacked image. Finally, for
each filter and each field, the individual exposures were combined
into a single deep one using \textsc{mscstack}.

\section{The galaxy sample}\label{sec:the_galaxy_sample}
Our galaxy sample comes from the four MOSAIC-II fields described
  in the previous section and is composed of 107 Antlia galaxies
considered as `members' and 77 new galaxies not catalogued before
\citep{2015MNRAS.451..791C}. The `member' galaxies were those selected
from the FS90 catalogue with membership status 1 (`definite members')
plus those which have measured radial velocities in the range of $1200
- 4200$\,km\,s$^{-1}$ \citep{2008MNRAS.386.2311S}. We can select
galaxies with membership status 1 from FS90 as `members' due to the
reliability of FS90 morphological membership classification, already
quantified in previous works \citep[e.g.][and references
  therein]{2012MNRAS.419.2472S}. Out of the 77 new galaxies, only 31
can be considered as `candidates' because they satisfy the following
criteria that ensure a reliable early-type morphological
classification: they have smooth and continuous profiles with
reasonable $S/N$ ratio out to the $\sim 27.5$ mag\,arcsec$^{-2}$ in
$R$-band, no obvious spiral structure present in the residuals of
the fits, etc. These criteria are fully explained in
\cite{2015MNRAS.451..791C}.  In addition, a photometric criterion was
also considered, from the colour-magnitude relation (CMR) for the
extended objects in the field: only new galaxies located within $\pm 3
\sigma$ of the CMR of the cluster members were selected. The CMR of
ETGs in galaxy clusters is a well defined, universal relation
with very small scatter \citep[e.g.][]{2008MNRAS.383..247P,
  2008AJ....135..380L, 2011MNRAS.410..280J, 2012AAS...21941106M}.

Our sample is $\sim$5 mag deeper than FS90, as the FS90
catalogue is complete to $B_{\mathrm{T}} = 18$\,mag, that corresponds
to $M_{B} = -14.7$\,mag at our adopted Antlia distance, while here we
reach $M_{T_1} \sim -12$\,mag, that corresponds to $M_{B} \sim
-9.6$\,mag \citep{1995PASP..107..945F}. 

\section{Isophotal analysis and computation of geometrical parameters}%
\label{sec:isophotal_analysis}
We used the \textsc{ellipse} \citep{1987MNRAS.226..747J} task within
the \textsc{isophote} IRAF's package, to obtain the observed surface
brightness profiles (surface brightness versus semi-major axis
$a$). The semi-major axis was transformed into equivalent radius ($r =
\sqrt{a~b} = a \sqrt{1-\epsilon}$, being $a$ the isophote semi-major
axis and $\epsilon$ the ellipticity) for all ETGs in the sample.

The initial values needed for the Fourier fitting, like the geometric
centre, initial ellipticity, and position angle of the first trial
ellipse, were estimated by visual inspection, for each galaxy in the
sample.  The intensity $I(\theta)$ along the trial ellipse is
described by a Fourier series,
\begin{equation}\label{eq:intensity}
  I(\theta) = \IO + \sum^{N}_{n=1} A_{n} \sin(n\theta) + B_{n}
  \cos(n\theta) \textnormal{,}
\end{equation} 
where $I_{0}$, is the mean isophotal intensity along the ellipse, $N$
is the highest harmonic fitted, $\theta$ is the azimuthal angle
measured from the major axis, and $A_{n}$ and $B_{n}$ with $n = 1, 2,
...$ are the harmonic amplitudes of the Fourier series. If the
isophotes were perfect ellipses (which is not the case for real
galaxies), the coefficients with $n \leq 2$ would be the only not null
ones. The fit begins with the assumption that the first two orders
($A_\mathrm{1}$, $A_\mathrm{2}$ , $B_\mathrm{1}$, $B_\mathrm{2}$) are
nonzero. The $A_{n}$ and $B_{n}$ coefficient provided by
\textsc{ellipse} are normalised to the semi-major axis $a$ and
corrected by the local intensity gradient.  The output ellipse
coefficients $B_{n}$ are converted to $a_{n}/a$ using
\begin{equation}\label{eq:coeff}
\frac{a_{n}}{a} = B_{n} \sqrt{1-\epsilon} = B_{n} \sqrt{b/a} \textnormal{.}
\end{equation} 

Once the parameters are obtained, the procedure continues with the
calculation of the third and fourth harmonic coefficients through a
least-squares fit. These coefficients ($A_\mathrm{3}$, $A_\mathrm{4}$,
$B_\mathrm{3}$ and $B_\mathrm{4}$) determine the deviation of the
isophote from a perfect ellipse.This procedure is repeated for the
next semi-major axis, defined by the variable $STEP$ in
\textsc{ellipse}, until it reaches a pre-defined value of the
semi-major axis. We used a linear step for each profile.  The
ellipticity and position angle are not well determined close to the
centre due to seeing; this effect will be analysed in
Section~\ref{sec:the_effect_of_seeing}.

The geometrical parameters, such as ellipticity and Fourier
coefficients, vary along the galactocentric radius of the
surface-brightness profile and, as a consequence, we cannot consider a
single characteristic value as representative of the entire galaxy. In
order to compare these parameters to other global galaxy properties,
we choose to estimate a weighted average value for each parameter along
different ranges of effective radius (\re). We divide each galaxy into
four regions: region 1, between the seeing radius (1\,arcsec) and
1.5\,\re; region 2, from 1.5 to 3.0\,\re; region 3, from 3.0 to
4.5\,\re; and region 4, further than 4.5\,\re.  Following
\cite{2014ApJ...787..102C}, we estimate each parameter within each
region by means of expressions like the following:
\begin{equation}\label{eq:mean_value}
 \left\langle \frac{a_{4}}{a} \right\rangle = \frac{ \int_{r_{s}}^{1.5
     \re} \frac{a_{4}(r)}{a} I(r) [\sigma_{\frac{a4}{a}}(r)]^{-2}
   \mathrm{d} r}{ \int_{r_{s}}^{1.5 \re} I(r)
   [\sigma_{\frac{a4}{a}}(r)]^{-2} \mathrm{d}r} \textnormal{,}
\end{equation}
which represents the mean weighted value of $a_{4}/a$ in region 1.
That is, all the calculated average parameters are weighted by
intensity (in counts) and inversely weighted by the corresponding
variance. Note that there will be fewer parameters assigned to region
4 because the fitting of the model to the profile is not always
reliable in the outer regions. Table~\ref{tab:geometric-parameter}
shows an example of the geometrical parameters computed for the
galaxies in the sample.

\begin{table*}
  \caption{Geometric parameters obtained for the galaxies in the
    sample. Columns: (1) id from FS90, (2)-(4) mean values calculated
    by Equation\,\ref{eq:mean_value} for \me, \mac and \mat on each
    radial range 1 to 4 (first to fourth line, when available). The
    full table is electronically
    available.}\label{tab:geometric-parameter}
\begin{center}
\begin{tabular}{c@{\hskip 24pt}c@{\hskip 24pt}c@{\hskip 24pt}c}
\hline
(1) & (2) & (3) & (4) \\
id (FS90) & \me & \mac & \mat \\
\hline
70    	&	 0.272$\pm$0.051 	&	 0.004$\pm$0.013 	&	 0.003$\pm$0.012 \\
	&	 0.330$\pm$0.013 	&	 -0.001$\pm$0.017 	&	 0.049$\pm$0.071 \\
	&	 0.306$\pm$0.001 	&	 0.031$\pm$0.016 	&	 0.027$\pm$0.016 \\
	&	 \nodata         	&	 \nodata        	&	 \nodata \\
72    	&	 0.334$\pm$0.062 	&	 -0.004$\pm$0.003 	&	 -0.003$\pm$0.002 \\
	&	 0.380$\pm$0.001 	&	 -0.002$\pm$0.001 	&	 0.002$\pm$0.005 \\
	&	 0.378$\pm$0.001 	&	 0.001$\pm$0.009 	&	 -0.008$\pm$0.004 \\
	&	 0.349$\pm$0.005 	&	 -0.010$\pm$0.034 	&	 0.006$\pm$0.024 \\
73    	&	 0.255$\pm$0.017 	&	 -0.001$\pm$0.004 	&	 -0.001$\pm$0.006 \\
	&	 0.245$\pm$0.018 	&	 -0.007$\pm$0.008 	&	 -0.005$\pm$0.015 \\
	&	 0.257$\pm$0.004 	&	 0.049$\pm$0.143 	&	 0.080$\pm$0.138 \\
	&	 \nodata         	&	 \nodata        	&	 \nodata \\
78    	&	 0.122$\pm$0.001 	&	 -0.012$\pm$0.051 	&	 -0.043$\pm$0.060 \\
        &	 \nodata                & 	 0.046$\pm$0.035 	&	 0.044$\pm$0.035 \\
	&	 \nodata        	&	 \nodata        	&	 \nodata \\
	&	 \nodata        	&	 \nodata        	&	 \nodata \\
79    	&	 0.301$\pm$0.090 	&        0.00$\pm$0.001         &        -0.002$\pm$0.002 \\
	&	 0.381$\pm$0.004 	&	 -0.001$\pm$0.002 	&	 0.00$\pm$0.003 \\
	&	 0.370$\pm$0.003 	&	 -0.002$\pm$0.008 	&	 0.008$\pm$0.004 \\
	&	 0.339$\pm$0.005 	&	 0.057$\pm$0.020 	&	 -0.030$\pm$0.025 \\
\hline
\end{tabular}
\end{center}
\end{table*}

\section{Surface Brightness Profiles}\label{sec:surface_brightness_profiles}
Given the large number of galaxies in the sample, and the fact that
the reduction procedure applied to obtain the surface-brightness
profiles consists of several steps that can be automatised (i.e. trim
the original image, estimate sky level around the galaxy, etc.), we
developed an IRAF pipeline in order to obtain results in an
homogeneous way. In this section, we describe such procedure adopted
to obtain the surface-brightness profiles.
  
The MOSAIC\,II images have 8800\,pixel $\times$ 8000\,pixel. Although
automatic detection software \citep[e.g.
SExtractor,][]{1996A&AS..117..393B} can be carefully configured for
faint sources identification, as we deal whith early type galaxies in
a nearby cluster, we decided to carry on the galaxy detection just by
visual inspection, which has been shown to be a very efficient method
in such a case. We started by re-identifying the FS90 galaxies and
then looked for new galaxies. After each galaxy detection, a subimage
of about 500 pixels $\times$ 500 pixels (135\,arcsec $\times$
135\,arcsec), centred on each object, was cut. Due to the large
MOSAIC\,II field, we preferred to estimate the background (sky level)
for each galaxy independently, instead of setting the same background
level for the whole image. The adopted size of these subimages was
large enough to make a good sky estimation. We first calculated an
initial value of the sky level taking the `mode' from several
positions around the galaxy, free from other sources, using the
\textsc{imexamine} task. Then, we subtracted that constant intensity
from the subimage and, due to the large-scale residual removal applied
on the previous reduction process, we found that our method to
estimate the sky was appropriate for the brightness level of the
sample. Once the calibrated galaxy profile was obtained, we applied an
iterative process to perform a second order correction of the sky
level, until the outer part of the integrated flux profile became as
flat as possible for large galactocentric distances.  Such corrections
remained between $\pm$\,10\,ADU (i.e, less than 5\% of the mean sky
level).

The last step before performing the fit of the model profile, was to
build a mask for each subimage to remove any objects that might have
affected the brightness profile, like foreground stars
  and cosmetics. In this way, we obtained one mask for each subimage
and each filter, using the \textsc{badpiximage} task.  We also took
into account objects hidden in the galaxy brightness, using different
display levels.  As a consequence, the final masks resulted more
accurate than those generated directly by the \textsc{ellipse} task.

Afterwards, we performed a first run of the \textsc{ellipse} task,
leaving all the geometric parameters `free', just to obtain
approximate values for the following initial geometric parameters:
\begin{enumerate}
\item \textsc{x0}, \textsc{y0}: coordinates of the initial isophote centre.
\item \textsc{pa0}: initial position angle.
\item \textsc{ellip0}: initial ellipticity.
\item \textsc{sma0}: initial semi-major axis length.
\item \textsc{maxsma}: maximum semi-major axis length.
\end{enumerate}

For each galaxy, we defined a set of initial parameters in such a way
to improve the stability of the isophotal fit. The minimum semi-major
axis (\textsc{minsma}) was taken as small as possible to be able to
fit the very central region of the galaxy. As the images were
sky-subtracted, we defined the value of \textsc{maxsma} as that for
which the galaxy brightness approaches zero level.  This procedure was
applied on the $R$ images as they are deeper than the $C$ ones
\citep{2015MNRAS.451..791C}. The $R$-band output table was later used
as input to \textsc{ellipse} on the $C$-band images to perform the
photometry.

If the image had defects that could complicate the fit, and/or the
galaxy was so faint that the brightness profile was strongly dependent
on the choice of the initial parameters, we kept one of them fixed to
allow for a better solution with less degrees of freedom. These
galaxies were mainly the faintest dwarf ellipticals (dE) or dwarf
spheroidals (dSph). Fixing one or more parameters in the iteration,
does not modify the total magnitude of the galaxy although information
on geometrical parameters may be lost.

\begin{figure*}
  \subfloat[]{%
    \label{fig:fs90211}\centering
    \includegraphics[width=0.9\columnwidth]{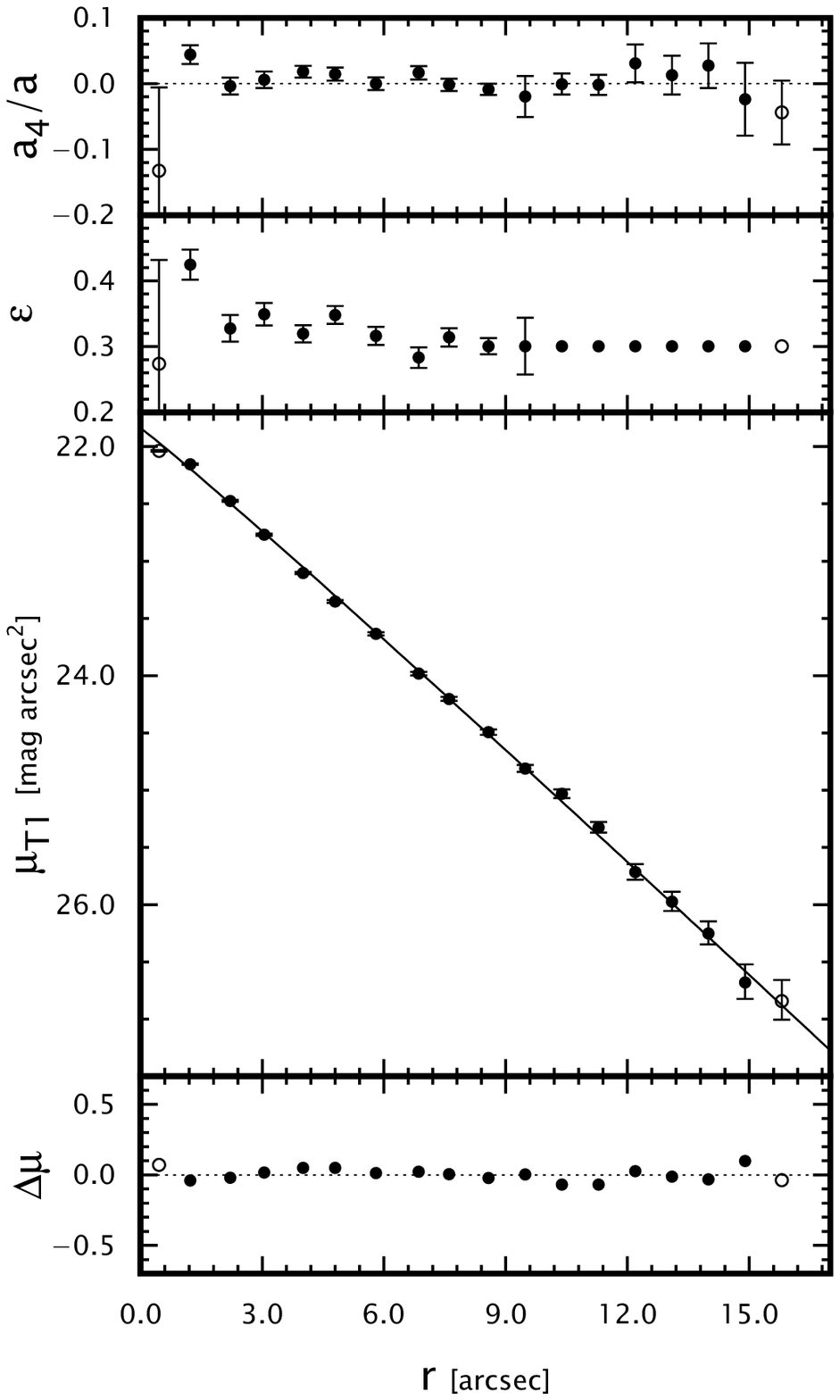}}\quad
  \subfloat[]{%
    \label{fig:fs90307}\centering
    \includegraphics[width=0.9\columnwidth]{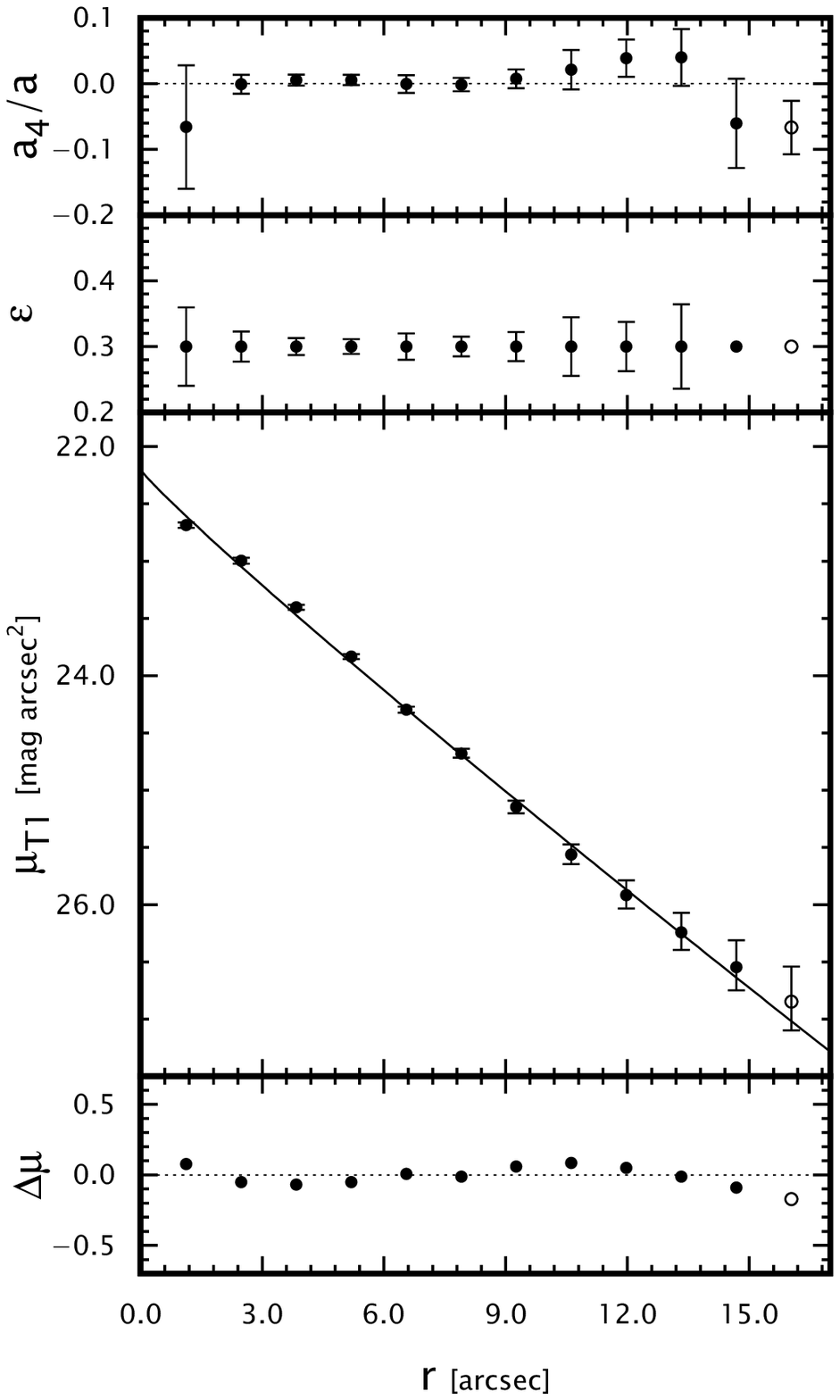}}\\
  \subfloat[]{%
    \label{fig:fs90211_1}\centering
    \includegraphics[width=0.9\columnwidth]{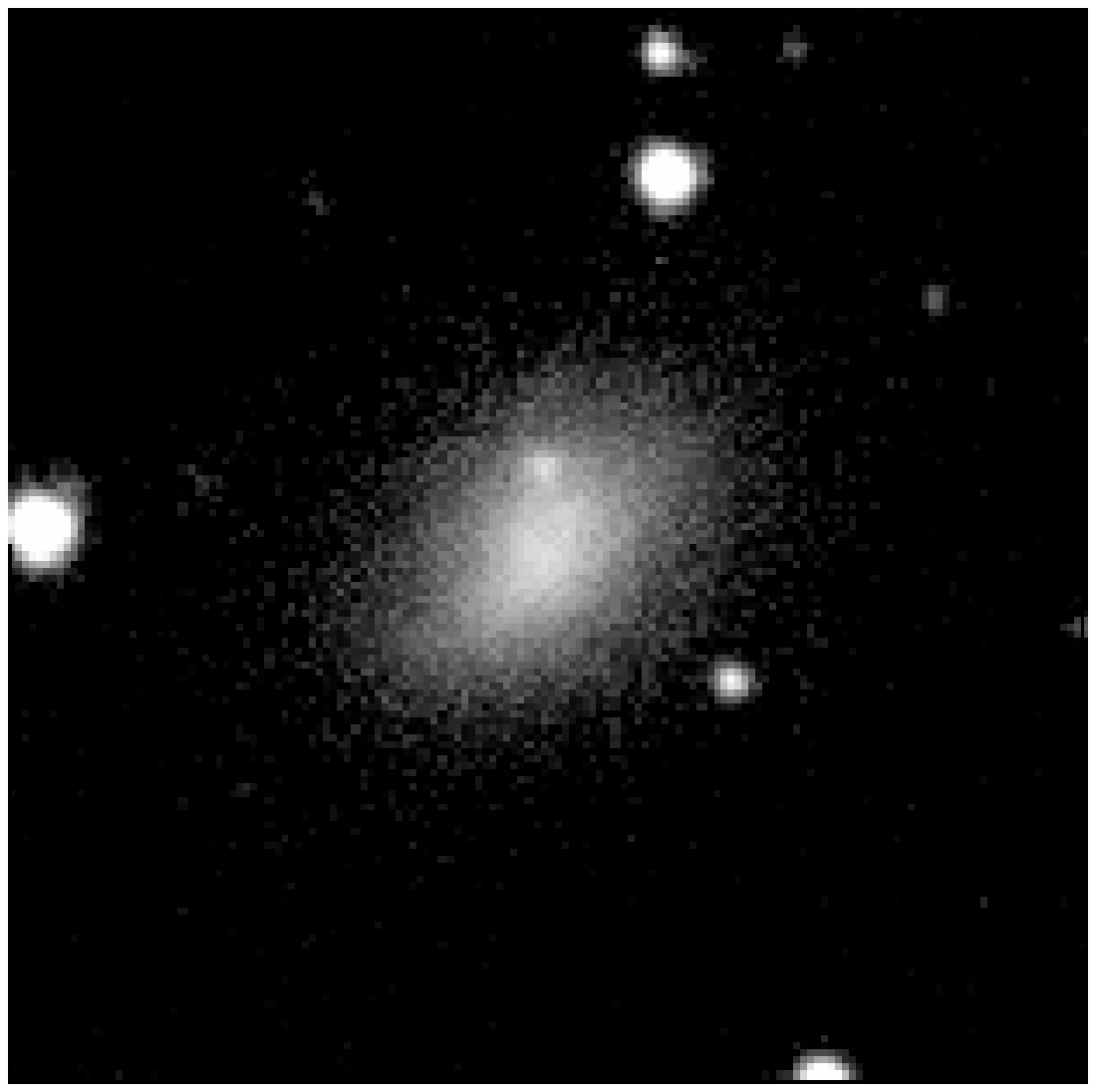}}\quad
  \subfloat[]{%
    \label{fig:fs90307_1}\centering
    \includegraphics[width=0.9\columnwidth]{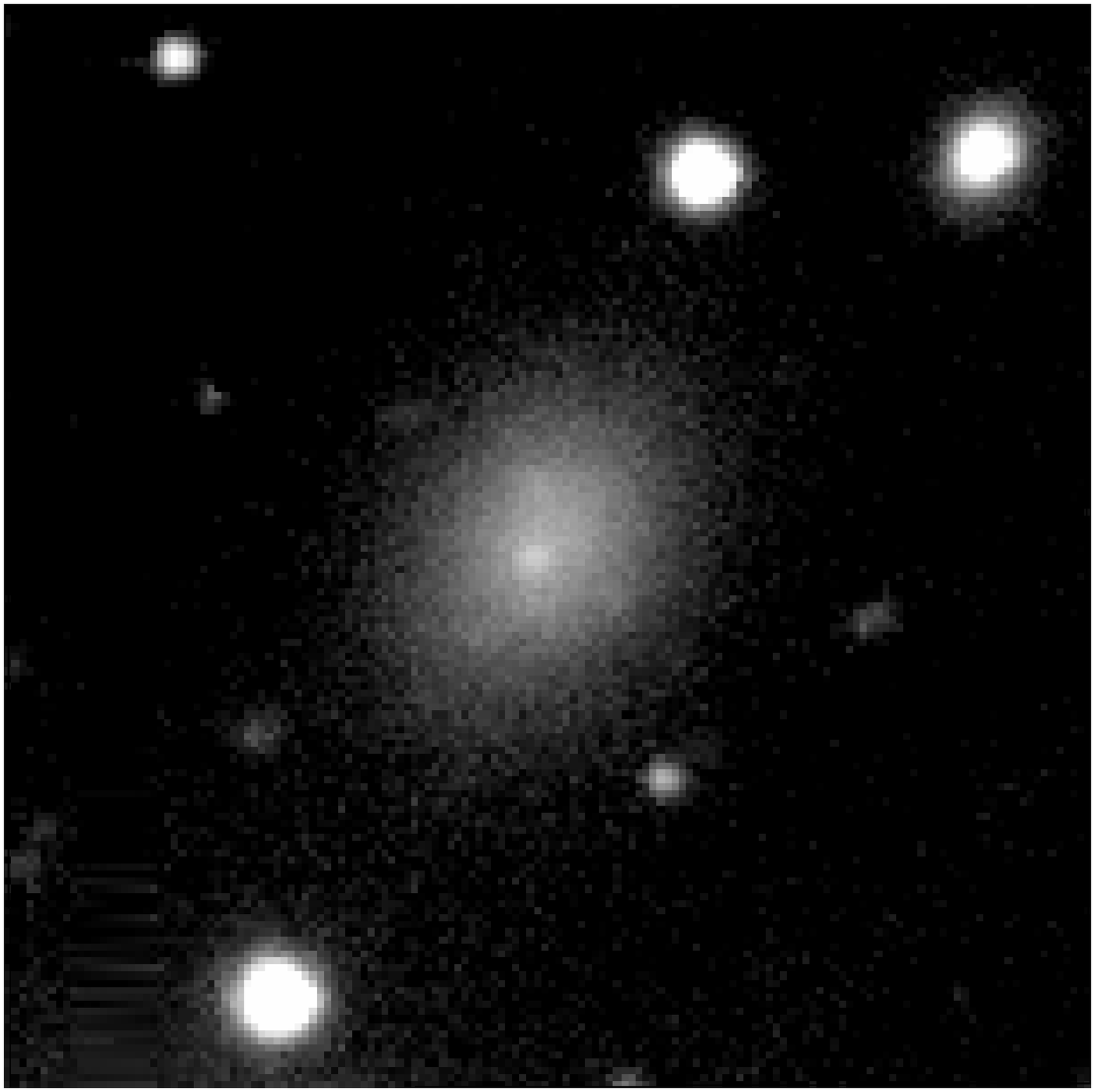}}
  \caption{Two examples of the profile fits on the $T_{1}$-band
    images: FS90\,211 (left) and FS90\,307 (right). Figures~(a) and
    (b) show (from top to bottom) the variation along $r$ of
    $a_{4}/a$, ellipticity ($\epsilon$), surface brightness
    ($\mu_\mathrm{T1}$) and the residual between the model and the
    observed profile. On each panel, we show with filled circles the
    data used in the fits and with open circles the discarded data,
    all of them with their respective error bars. Figures~(c) and (d)
    show $T_{1}$-band images of each galaxy (40 $\times$ 40
    arcsec).}\label{fig:profile-examples}
\end{figure*}

Figure~\ref{fig:profile-examples} shows examples of the galaxy
brightness profiles of two galaxies in the sample (FS90\,211 on the
left and FS90\,307 on the right). From top to bottom, the figure
presents the run along $r$ of: $a_{4}/a$, $\epsilon$, surface
brightness $\mu_{T_1}$ (filled circles) along with the fitted S\'ersic
model (continuous line), and the corresponding residual between model
and observed profile. Fnally, the $T_{1}$-band image is shown.

\subsection{Numerical fits to the surface brightness profiles}%
\label{sec:numerical_fit_profile}
To fit the brightness profiles, we used an uni-dimensional S\'ersic model
\citep{1968adga.book.....S}, which can be expressed as follows
\begin{equation}\label{eq:modelo_de_sersic_intensidades}
  I(r) = \Ie \cdot e^{-b_{n} \left[
      \left(\frac{r}{\re}\right)^{1/n} - 1
      \right]}\textnormal{,}
\end{equation} 
while in magnitudes per square arcsec it is:
\begin{equation}\label{eq:modelo_de_sersic}
  \mu(r) = \mue + 1.0857 \cdot b_{n} \left[
    \left(\frac{r}{\re}\right)^{1/n} - 1
    \right]\textnormal{,}
\end{equation} 
where \re is the effective radius, \mue is the effective surface
brightness, and $n$ is the S\'ersic index, which is a measure of the
concentration of the profile. The constant $b_{n}$ depends on the
shape parameter $n$ and is obtained numerically by solving the
following equation \citep{1991A&A...249...99C},
\begin{equation}\label{eq:bene}
\frac{\Gamma(2n)}{2} = \gamma(2n,b_{n})\textnormal{,}
\end{equation} 
where $\Gamma(x)$ is the complete gamma function and $\gamma(a, x)$
the incomplete gamma function. An alternative way to express the
S\'ersic model, in terms of intensity, is the following:
\begin{equation}\label{eq:modelo_de_sersic_simplificado}
  I(r) = \IO e^{-\left(\frac{r}{\rO}\right)^{N}}\textnormal{,}
\end{equation} 
where \IO is the central surface brightness, \rO is a
scale parameter and $N = 1/n$. If we express the above equation in
units of magnitude per square arcsec:
\begin{equation}\label{eq:modelo_de_sersic_magnitudes}
  \mu(r) = \muO + 1.0857
  \left(\frac{r}{\rO}\right)^{N}\textnormal{;}
\end{equation} 
which is the one used for our profile fits, where \muO is the central
surface brightness. The transformation between effective radius and
scale parameter can be obtained using equations
\ref{eq:modelo_de_sersic_intensidades} and
\ref{eq:modelo_de_sersic_magnitudes}:
\begin{equation}\label{eq:sersic_transformacion}
\IO e^{-\left(\frac{r}{\rO}\right)^{N}} = \Ie e^{-b_{n}
  \left[ \left(\frac{r}{\re}\right)^{1/n} - 1 \right]} =
e^{b_{n}} \Ie e^{-b_{n}
  \left(\frac{r}{\re}\right)^{1/n}}\textnormal{.}
\end{equation}
Considering $r = 0$ we obtain, 
\begin{eqnarray}\label{eq:sersic_transformacion_1}
\IO &=& e^{b_{n}} \Ie \\
\textnormal{and then, } \rO &=& b_{n}^{-n} \re\textnormal{.}
\end{eqnarray} 
The total flux is obtained by solving the integral:
\begin{equation}\label{eq:sersic_flujo}
F_\mathrm{T} = \int_{0}^{\infty} 2 \pi I(r) r \,\mathrm{d}r = 2 \pi \int_{0}^{\infty}
e^{-b_{n} \left[ \left(\frac{r}{\re}\right)^{1/n} - 1
    \right]} r \,\mathrm{d}r \textnormal{,}
\end{equation} 
which leads us to:
\begin{equation}\label{eq:sersic_flujo_2}
F_\mathrm{T} = 2 \pi \re^{2} b_{n}^{-2n} \Ie n e^{b_{n}} n \Gamma(2n) \textnormal{.}
\end{equation}
The integral magnitude is obtained by transforming the above equation,
\begin{eqnarray}\label{eq:sersic_int}
m &=& C_\mathrm{0} - 2.5 \log\left(2 \pi \re^{2} b_{n}^{-2n} \Ie n e^{b_{n}} n \Gamma(2n)\right) \\
m &=& \muO - 1.99 - 5 \log(\re) + 5~n \log(b_n) - \\
  & &   -2.5\log\left(n~\Gamma(2n)\right) \nonumber\textnormal{.}
\end{eqnarray} 

The fits were obtained using the task \textsc{nfit1d} from
\textsc{IRAF}, which implements the $\chi^{2}$ statistic test through
the Levenberg-Marquardt algorithm. We excluded the inner arcsec of the
profile in the fits, in order to minimize {\it seeing} effects.  We
will show in the next sub-section that, in this way, the fits are not
significantly affected by seeing for galaxies with $n \lesssim 3$.  In
most cases, we have been able to fit the profiles with a single
S\'ersic model obtaining residuals smaller than $0.5$\,mag. We want to
remark that the scale parameters presented in this paper have been
derived without trying a bulge-disc profile
decomposition. Table~\ref{tab:small-catalog} shows some of the scaling
parameters and photometric magnitudes obtained for the sample.
\begin{table*}
\caption{Basic parameters of the Antlia galaxy sample: (1) id from
  FS90, (2)-(3) J2000 coordinates, (4) morphology from FS90, (5)
  Galactic extinction from \protect\cite{2011ApJ...737..103S},
  (6)-(12) global properties calculated in this work (S\'ersic index,
  central surface brightness, scale radius, effective surface
  brightness, effective radius, $T_{1}$-band magnitude, $(C-T_{1})$
  colour), (13) radial velocity. The full table can be accessed
  electronically.}
\label{tab:small-catalog}
\begin{tabular}{ccccccccccccc}
\hline
(1) & (2) & (3) & (4) &  (5) & (6) & (7) & (8) & (9) & (10) & (11) & (12) & (13)\\
FS90 & RA & DEC & FS90 & $E(B-V)$ & n & \muO & \rO  & \mue & \re & $T_{1}$ & $(C-T_{1})$ & $v_{r}$ \\
id &  J2000 & J2000 & morph. &  mag & -- & mag arcsec$^{-2}$ & arcsec & mag arcsec$^{-2}$ & arcsec & mag & mag & km seg$^{-1}$ \\
\hline
70    	&	 10:29:10 	&	 -35:35:20 	&	  dE  	&	 0.070 	&	  1.15 	&	 22.08 	&	  2.65 	&	 24.23 	&	  5.83 	&	 17.65 	&	  1.75 	& 2864 $\pm$ 70$^{\text{a}}$  \\
72    	&	 10:29:20 	&	 -35:38:24 	&	  S0  	&	 0.067 	&	  1.58 	&	 18.11 	&	  1.19 	&	 21.18 	&	  6.14 	&	 14.33 	&	  1.93 	& 2986 $\pm$ 38$^{\text{b}}$  \\
73    	&	 10:28:10 	&	 -35:42:55 	&	  dE  	&	 0.065 	&	  1.28 	&	 20.54 	&	  1.57 	&	 22.96 	&	  4.38 	&	 16.95 	&	  1.69 	&  \nodata  		 \\
78    	&	 10:28:16 	&	 -35:46:24 	&	  dE  	&	 0.067 	&	  0.78 	&	 23.71 	&	  4.42 	&	 25.07 	&	  5.27 	&	 18.88 	&	  1.68 	&  \nodata  		 \\
79    	&	 10:28:19 	&	 -35:27:16 	&	  S0  	&	 0.073 	&	  1.86 	&	 16.73 	&	  0.71 	&	 20.43 	&	  6.98 	&	 13.21 	&	  2.12 	& 2734 $\pm$ 36$^{\text{b}}$  \\
80    	&	 10:28:19 	&	 -35:45:30 	&	  dS0  	&	 0.066 	&	  2.27 	&	 15.87 	&	  0.20 	&	 20.43 	&	  5.17 	&	 13.77 	&	  2.10 	& 2519 $\pm$ 31$^{\text{b}}$  \\
84    	&	 10:28:23 	&	 -35:31:46 	&	  E  	&	 0.073 	&	  1.89 	&	 16.65 	&	  0.53 	&	 20.39 	&	  5.50 	&	 13.69 	&	  2.05 	& 2428 $\pm$ 30$^{\text{b}}$  \\
85    	&	 10:28:24 	&	 -35:34:21 	&	  dE  	&	 0.072 	&	  0.61 	&	 23.22 	&	  4.44 	&	 24.20 	&	  4.18 	&	 18.62 	&	  1.68 	& 2000 $\pm$ 200$^{\text{a}}$  \\
94    	&	 10:28:31 	&	 -35:42:18 	&	  S0  	&	 0.065 	&	  2.60 	&	 13.72 	&	  0.06 	&	 19.01 	&	  3.37 	&	 13.20 	&	  2.01 	& 2786 $\pm$ 45$^{\text{b}}$  \\
103    	&	 10:28:45 	&	 -35:34:38 	&	  dE  	&	 0.075 	&	  2.46 	&	 20.75 	&	  0.11 	&	 25.73 	&	  4.83 	&	 19.18 	&	  1.98 	& 2092 $\pm$ 29$^{\text{b}}$  \\
\hline
\end{tabular}
\begin{flushleft}%
  Notes.- Radial velocities are from: a=\cite{2012MNRAS.419.2472S},
  b=NED\footnotemark, c=\cite{2015A&A...584A.125C}.%
\end{flushleft}%
\end{table*}
\footnotetext{This research has made use of the NASA/IPAC
  Extragalactic Database (NED).}

\subsection{Effects of {\it seeing} on the modelled parameters}%
\label{sec:the_effect_of_seeing}
Ground based images are affected by atmospheric seeing; for images of
extended objects it always acts distributing light from higher- to
lower-surface brightness regions, thus mainly affecting the central
portions of early-type galaxies profiles.

Instead of modelling out seeing effects on the fitted
parameters \citep{2001MNRAS.321..269T, 2001MNRAS.328..977T}, we
performed a series of simple simulations of artificial galaxies
following the procedure explained by \cite{2005A&A...430..411G}.
Using the task \textsc{mkobjects} from IRAF \textsc{artdata} package,
we built a series of FITS images of simulated galaxies with S\'ersic
light profiles and null ellipticity.  In addition, we fixed \muO
in 10\,\mags, while the S\'ersic index ranged between 0.5 to
4. Finally, we added to each simulated image, a sky level and noise
similar to those on the real images.

To simulate the seeing effects, we performed a convolution using the
\textsc{gauss} task, with Gaussian profiles and $\sigma$ ($\sigma =
0.42466$\,FWHM) in the range of 0.5-10\,arcsec. The S\'ersic model was
fitted to the simulated galaxies following exactly the same procedure
as for the real galaxies, excluding the innermost arcsec from the
profile.
\begin{figure}
  \centering
  \includegraphics[width=0.85\columnwidth]{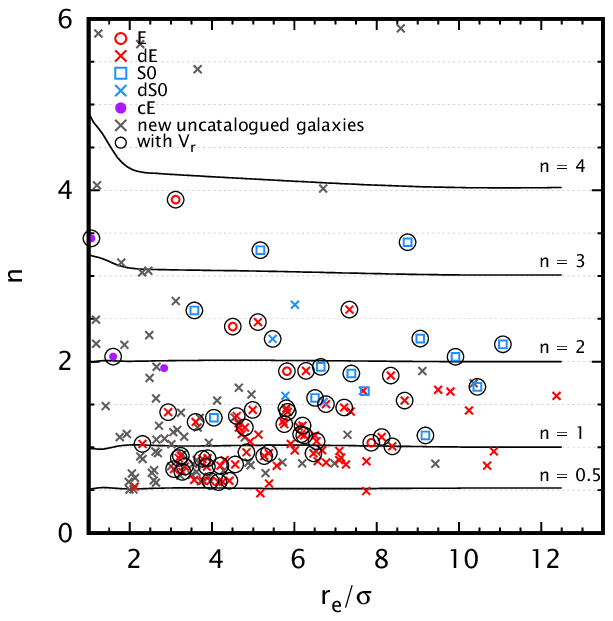}
  \caption{Measured S\'ersic index $n$ versus effective radius \re (in
    units of the PSF dispersion).  The solid lines show the results
    for artificial galaxies and the symbols for real
    ones. }\label{fig:re_n_sim}
\end{figure}

Figure~\ref{fig:re_n_sim} shows the results obtained for S\'ersic
index $n$ versus \re/$\sigma$. The symbols indicate different galaxy
morphologies and the lines correspond to different theoretical (model)
S\'ersic indices.  It can be seen that for small $n$ values, the
S\'ersic indices measured from the convolved fake galaxies follow
reasonably well their theoretical values; however, for $n > 3$ there
are significant differences for small \re/$\sigma$, in the sense that
the measured $n$ is overestimated. This result is in agreement with
that obtained by \cite{2005A&A...430..411G}, and it is due to light
being distributed off the galaxy centre by seeing, thus leading to a
measured S\'ersic index that is higher than the intrinsic one. This
effect is of course stronger for more concentrated (i.e., $n > 3$)
profiles.

Given that there are very few real galaxies in the sample
within the ranges of $n$ and \re where the effect of seeing is
significant, it was decided not to perform a general correction for
seeing.

\subsection{Signal-to-noise ratio}\label{sec:signal-to-noise}
In order to estimate the quality of the profile fits and the
parameters obtained, we calculated how the signal-to-noise varies as a
function of the equivalent radius of the profile using the following
expression \citep{2011MNRAS.414.2055M}:
\begin{equation}\label{eq:sn}
  \frac{S}{N}(r) = \frac{I_{t}(r)\,{\scriptstyle
      [\textnormal{pixel}^{-2}]} \cdot \sqrt{A_\mathrm{iso}}\,{\scriptstyle
      [\textnormal{pixel}]}}{\sqrt{I_\mathrm{s}}\,{\scriptstyle
      [\textnormal{pixel}^{-1}]}}\textnormal{,}
\end{equation}
where $A_\mathrm{iso}$ is the area of the given isophote in pixels$^{2}$,
\begin{equation}\label{eq:sn_A}
  A_\mathrm{iso} = 2\pi \sqrt{0.5 (a^{2} + b^{2})} \textnormal{,}
\end{equation}
$a$ and $b$ are the semi-axes (major and minor) of the elliptic
isophote, $I_{t}(r)$ the total surface brightness of the galaxy and
$I_\mathrm{s}$ the sky surface brightness. The $S/N$ ratios for the
fainter galaxies in the present sample ($T_{1} > 14$\,mag) at the
isophote 27.5\,\mags range between 1.6 $\pm$ 0.3 to 3.0 $\pm$ 1.0.

\section{Results}\label{sec:results}%
\subsection{Comparison between isophotal and model-fit effective radii}
\begin{figure}
  \subfloat[Difference between \re calculated by
    \protect\cite{2015MNRAS.451..791C} fitting a S\'ersic model and the
    `isophotal' $\sim 27$ mag\,arcsec$^{-2}$ ones, versus absolute and
    apparent $T_1$ magnitudes (top and bottom axis),
    respectively.]{%
    \label{fig:Mv_re}%
    \includegraphics[width=0.95\columnwidth]{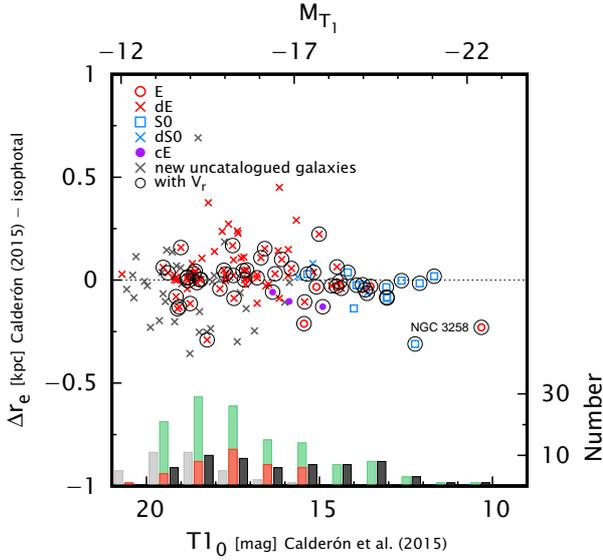}}\\
  \subfloat[Same as plot (a) but for the difference in effective surface
    brightness.]{%
    \label{fig:Mv_mue}%
    \includegraphics[width=0.95\columnwidth]{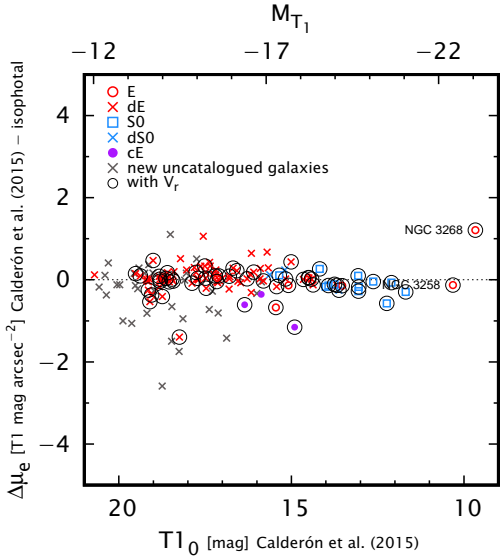}}
  \caption{Differences between structural parameters obtained from
    S\'ersic fits and isophotal ones.}\label{fig:scaling_relations}%
\end{figure}

The effective radius may be measured in different ways. In our
case, we obtained it by fitting a single S\'ersic model to the
observed profile so that \re encloses half the
luminosity of the model integrated to infinity
\citep{2015MNRAS.451..791C}. 
Now, we want to compare these effective radii with the ones calculated
directly from the isophotal parameters corresponding to $\sim 27$
mag\,arcsec$^{-2}$ in the $T_1$ band. Figure~\ref{fig:Mv_re} shows the
difference between the \re calculated by \cite{2015MNRAS.451..791C}
performing an extrapolation to infinity and the `isophotal' ones,
versus absolute (top axis) and apparent (bottom axis) $T_{1}$
magnitudes.  At the bottom of the same Figure, we include an histogram
that shows the number of galaxies considered in each magnitude bin,
depicted on the right axis. The total galaxy sample is represented in
green, candidate members in light grey, members in red, and members
with measured radial velocity in black.

It is important to remark that for the two brightest galaxies
($M_{T_1} < -22$\,mag), the effective radius results underestimated
when using a single component profile (for NGC\,3268 the difference is
even larger than 1\,kpc). A similar (although milder) tendency is
present for S0s and cEs.  It can be seen that, as expected, the
confirmed dEs show mostly positive differences, while the new galaxies
(mainly candidates) are the ones showing more negative
differences. This effect is less noticeable if we consider a similar
difference but for the effective surface brightness
(Figure~\ref{fig:Mv_mue}). In this case, the confirmed dEs are evenly
distributed about zero.

\subsection{Geometrical parameters at different galactocentric radii}
\begin{figure}
  \begin{center}
    \includegraphics[width=0.95\columnwidth]{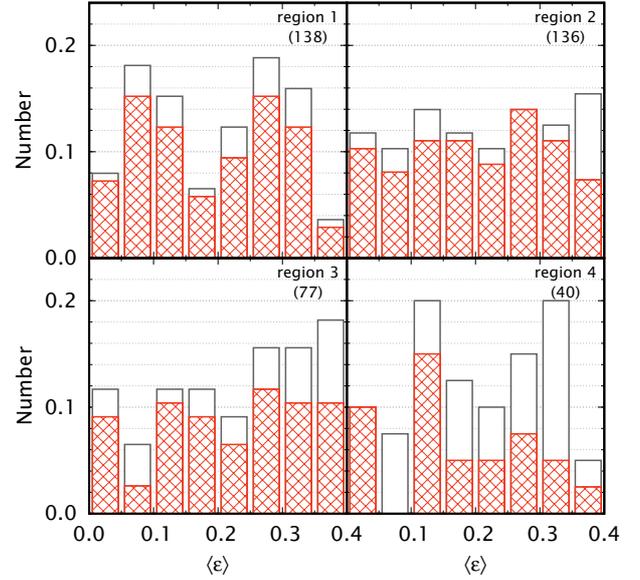}%
    \caption{Histograms of the mean weighted ellipticity
      distributions, cross-hatched in red for dE and dSph and open for
      the whole sample. Number of galaxies in each region are
      indicated in parentheses.}\label{fig:hist_e-multia}%
  \end{center}
\end{figure}

In this Section, we analyse the geometrical parameters obtained from
the \textsc{ellipse} output for the whole sample, considering the four
radial ranges (regions 1 to 4) defined in
Section~\ref{sec:isophotal_analysis}. Figure~\ref{fig:hist_e-multia}
shows the distribution of the intensity-weighted average ellipticity
\me for the four regions, with a cross-hatched (red) histogram for 
faint galaxies (dEs and dSph) and an open one 
for the whole sample. We note that the morphological
classification was done by visual inspection of each galaxy, following
the criteria used in FS90. That is why we do not establish a magnitude
limit (usually set around $M_{B} = -18$\,mag); an overlap in
luminosity between bright and dwarf ellipticals can thus be seen.

The histograms of mean ellipticity show flatter (although slightly
less extended) distributions, as compared to those obtained by
\cite{2014ApJ...787..102C} and \cite{2006MNRAS.370.1339H}, where a
main peak at \me $\sim 0.1 \to 0.16$ is evident in regions 1 to 4,
implying a dominant fraction of nearly round galaxies. Besides a
similar low \me peak, a second peak at \me $\approx 0.3$ is also
evident in region 1 of our sample. This reflects the fact that most of
the brighter galaxies in Antlia are lenticulars (S0), while dwarfs
also tend to exhibit relatively large flattenings, despite most of
them being classified as dE (dS0s are found only among the brighter
dwarfs).

\begin{figure}
  \begin{center}
    \includegraphics[width=0.95\columnwidth]{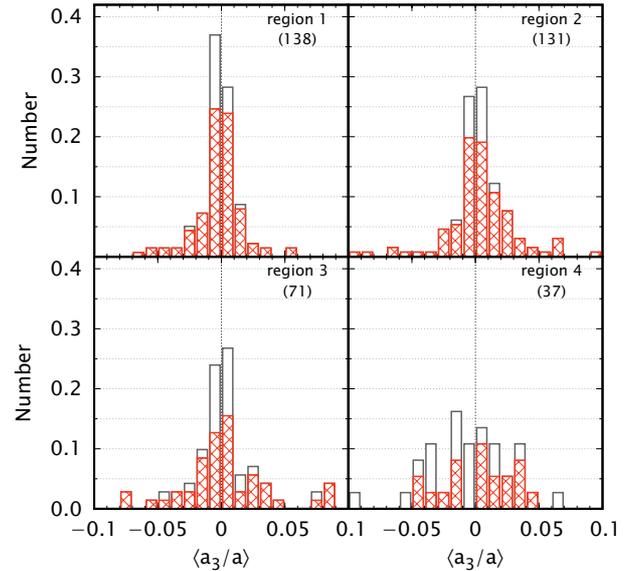}
    \caption{Same as Figure~\ref{fig:hist_e-multia} but for the \mat
      parameter distributions.}\label{fig:hist_a3-multia}%
  \end{center}
\end{figure}

\begin{figure}
  \begin{center}
    \includegraphics[width=0.95\columnwidth]{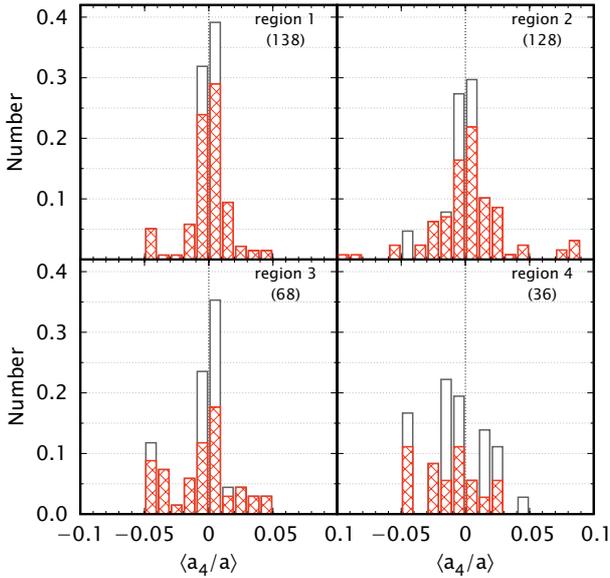}
     \caption{Same as Figure~\ref{fig:hist_e-multia} but for the \mac
       parameter distributions.}\label{fig:hist_a4-multia}%
  \end{center}
\end{figure}

Figures~\ref{fig:hist_a3-multia} and \ref{fig:hist_a4-multia} show
the distributions of the weighted mean values of the Fourier
coefficients \mat and \mac. As usual, they are reasonably fitted with
a single Gaussian centred at zero, except in the outermost
region, where the distribution is much flatter. In particular,
the coefficient \mac is slightly positive in regions 1 through 3 for
the dwarf galaxies, which indicates an excess of discy isophotes.  On
the contrary, the brighter galaxies show an excess of negative values
in region 2, pointing to boxy
isophotes. Figure~\ref{fig:hist_a3-multia} shows, for the brighter
galaxies, an excess of negative values of the \mat coefficient in the
innermost region; this can be related with minor mergers
\citep{2001MNRAS.326.1141R}.

\begin{figure}
  \subfloat[Region 2 versus region 1.]{\label{fig:e_e-1}%
    \includegraphics[width=0.45\columnwidth]{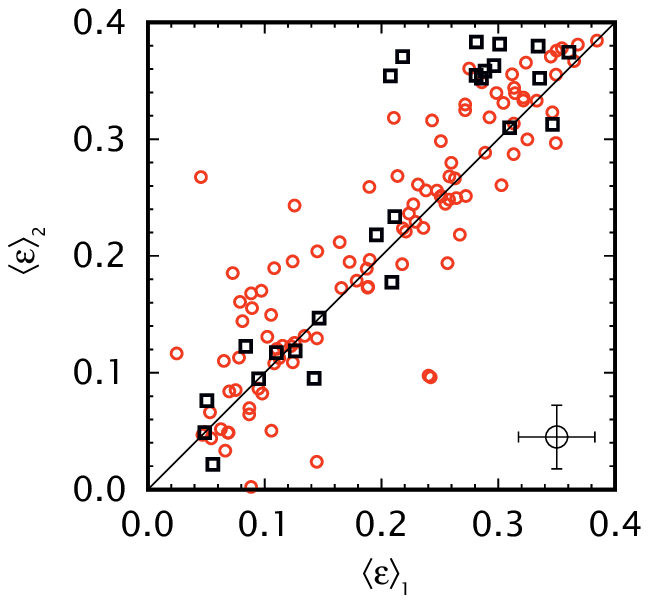}}\quad
  \subfloat[Region 3 versus region 2.]{\label{fig:e_e-2}%
    \includegraphics[width=0.45\columnwidth]{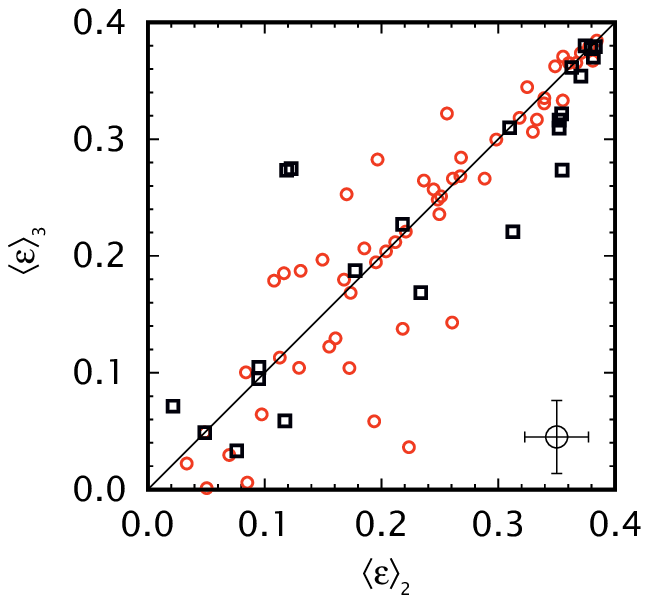}}\\
  \subfloat[Region 4 versus region 3.]{\label{fig:e_e-3}%
    \includegraphics[width=0.45\columnwidth]{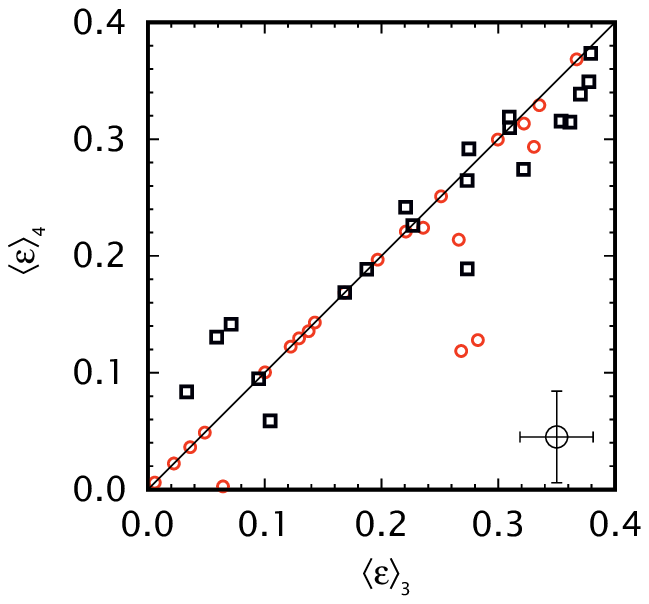}}%
  \caption{Comparison of \me values between adjacent regions. The open
    circle at the lower right corner shows the median error bars for
    each panel.}\label{fig:e_regions}%
\end{figure}

In order to assess the significance of any differences between the
weighted-mean values along the equivalent radius, we perform a
two-sample Kolmogorov-Smirnov (KS) test between adjacent regions
\citep{1992nrca.book.....P}, considering the whole sample. Regarding
the mean ellipticity (Figure~\ref{fig:e_regions}), the KS test shows
that adjacent regions may share the same distribution (see
Table~\ref{tab:kstest-e}). The two-peaked distribution for region 1,
although visually evident in Figure~\ref{fig:hist_e-multia}, is thus
not significantly different (at a 5\% significance level) from the
distributions in the other regions. With this in mind, we plot
separately with red circles the fainter galaxies of the sample and
with back squares the brighter sample, to compare their respective
behaviours. While dwarfs seem to be mostly responsible for the
disappearance of the \me $\approx 0.1$ peak when going from region 1
to region 2, brighter galaxies play this role for the \me $\approx
0.3$ peak. A qualitative analysis of Figure~\ref{fig:e_regions}, then,
shows that dwarfs on the low-$\epsilon$ peak in region 1 have been
shifted to both higher and lower ellipticities in region 2, while
bright galaxies on the high-$\epsilon$ peak have been mostly shifted
to still higher ellipticities. This means that some of the brighter
galaxies display positive ellipticity gradients (consistent with a S0
morphology), while dwarf galaxies may display either positive or
negative gradients.
\begin{table}
\caption{Results from the two-sample Kolmogorov-Smirnov test
  ($D$) for the mean ellipticity. The probability to support the
  hypothesis that the compared distributions are taken from the same
  parent distribution is given by $P$.}\label{tab:kstest-e}%
\begin{center}
{\renewcommand{\arraystretch}{1.1}
\begin{tabular}{c|c|c|c|c|c|c}
\hline
  & \multicolumn{2}{c}{Region 2} & \multicolumn{2}{c}{Region 3} &%
\multicolumn{2}{c}{Region 4} \\ 
         &  $D$  &  $P$  &  $D$  &  $P$  &  $D$  &  $P$  \\ \hline
Region 1 & 0.125 & 0.222 & 0.166 & 0.118 & 0.130 & 0.634 \\ 
Region 2 &       &       & 0.108 & 0.579 & 0.120 & 0.732 \\ 
Region 3 &       &       &       &       & 0.148 & 0.570 \\ \hline
\end{tabular}}
\end{center}
\end{table}
 
\subsection{Relations between \mac and \me, S\'ersic
  index and luminosity}%
\begin{figure*}
  \subfloat[]{\label{fig:e_a4}%
    \includegraphics[width=0.66\columnwidth]{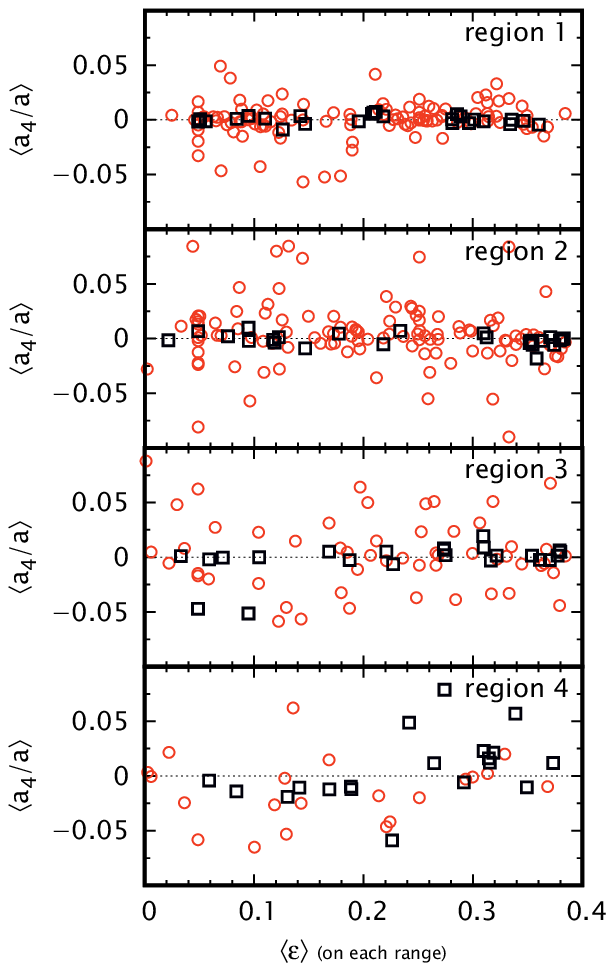}}\quad%
  \subfloat[]{\label{fig:n_a4}%
    \includegraphics[width=0.66\columnwidth]{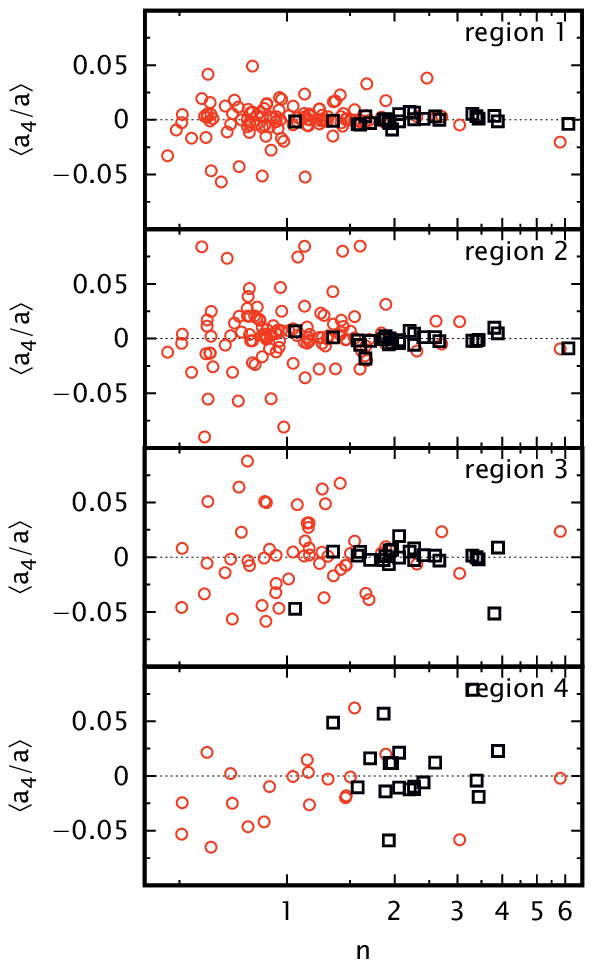}}\quad%
  \subfloat[]{\label{fig:M_a4}%
    \includegraphics[width=0.66\columnwidth]{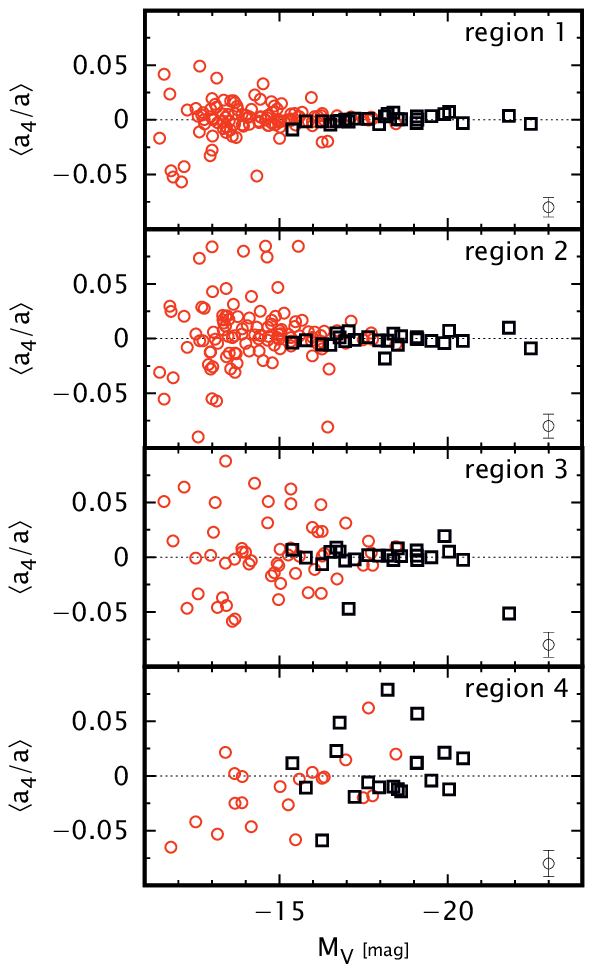}}%
  \caption{Relation between \mac and (a) \me, (b) $n$, and (c)
    luminosity. Each panel, on each figure, represents regions 1 to 4,
    from top to bottom. We identify dE and dSph galaxies of the
    sample with red circles and the brighter ones with black squares. 
    The error bars on the last
    figure are as in Figure~\ref{fig:e_regions}.}\label{fig:e_n_M-a4}%
\end{figure*}

Figure~\ref{fig:e_a4} shows the relation between \mac and ellipticity
(on each region).  As already shown in Figure~\ref{fig:hist_a4-multia},
we see that the \mac distributions get broader for larger radii
(regions 1 to 4), with a slight excess of positive \mac values in the
first two regions, indicating a dominance of discy isophotes. These
features are evident along the full range of ellipticities, so there
is no trend of \mac with \me. Figure~\ref{fig:n_a4} shows no clear
correlation between \mac and S\'ersic index. There is a group of
galaxies with negative \mac and low $n$ in region 1, but the tendency
is washed out in the outermost regions. Finally, Figure~\ref{fig:M_a4}
clearly shows that the dispersion in the \mac distribution increases
with decreasing galaxy luminosity, with an important increment for
galaxies with $M_{T_1} > -16$\,mag.  The tendency in region 1 is not
clear; however, in regions 2 and 3 there are more galaxies
(particularly dwarfs) with discy isophotes. As in the other plots, the
scatter of the relation increases rapidly as we go through region 1 to
region 4. These plots are in agreement with
\cite{2014ApJ...787..102C}, extending the range of surface
brightnesses at the faint end.

We applied the Spearman's rank correlation ($\rho$) test
\citep{spearman04}, which is used to decide whether a pair of
variables are correlated or not, to the data of
Figure~\ref{fig:e_n_M-a4}. Its advantages over the Pearson correlation
test are that it is non-parametric, and a linear relationship between
the variables is not a requirement. When we consider the complete
sample, the test results in small values for the \mac--$n$ relation
in regions 1 and 4, which implies a high correlation probability
between both variables. The Spearman coefficients are $\rho = 0.09$
and $\rho = 0.17$, respectively, which lead to probabilities $p =
0.24$ and $p = 0.28$ that the null hypothesis (i.e. no correlation) is
true.  Almost the same happens if we just consider the dE and dSph
galaxies. For the relation \mac versus \me the picture is similar,
although just considering regions 2 and 4.  The correlation
coefficient values given by the test are: $\rho = -0.2$ for region 2,
and $\rho = 0.4$ for region 4; the probability of the null hypothesis
(no correlation) being true is $p = 0.01$ for both regions.

\section{Discussion}\label{sec:discussion}
The study of the distributions of isophotal parameters in different
ranges along the radial profiles of the galaxies, is used as a tool to
look for statistical differences between inner and outer parts of the
galaxies, and their possible correlations with global galaxy
properties. In this section, we compare our results with
  numerical simulations that involve galaxy mergers that took place
  out of any deep gravitational potential, such as a cluster.  Thus,
  it should be taken into account that repeated tidal interactions may
  further affect the structural properties of cluster galaxies. Also,
  the merges themselves can be modified by the cluster potential
  well.
 
The ellipticity distribution (Figure~\ref{fig:hist_e-multia}) in our
sample shows a main peak around \me $\sim 0.28$, and a second peak
around \me $\sim 0.08$; this makes the galaxies in the Antlia cluster
more flattened in comparison to the samples presented in
\cite{2014ApJ...787..102C} and \cite{2006MNRAS.370.1339H}. These
differences may be explained by the distinctive characteristics of the
galaxy sample, the Antlia one being dominated by lenticular galaxies.
The shapes of the ellipticity histograms in regions 1 and 4 are
similar, showing two peaks around the same mean ellipticities. This
still holds when we consider the full range of radii. The dEs, which
are shown in red, follow the same trend as all the other morphologies;
this is true for all regions, except for region 4, which shows a large
fraction of rounder dEs. This behaviour is also found in hydrodynamic
simulations \citep{2015MNRAS.453..469T}. The KS test, however, shows
no statistical differences between the ellipticity distributions for
the inner and outer regions of the galaxies in the sample
(Table~\ref{tab:kstest-e}), so the above mentioned differences should
be regarded as marginal.

The deviations from perfect ellipses, measured by the Fourier
coefficients, have been studied since
\cite{1985MNRAS.216..429L}. However, several issues are pending and a
new discussion is still relevant. The $a_{4}$ coefficient is an
intrinsic parameter of the galaxy, without (projection) dependence on
the viewing angle. \cite{2005MNRAS.359.1379K} used $N$-body
simulations to predict that the percentage of discy-boxy galaxies is
affected by the environment, so that in overdense regions, galaxies
with more discy shape isophotes are produced \citep[see
also][]{2006ApJ...636..115P}.

The separation in radial bins of \mac shows that, for our sample, the
distributions of the two innermost regions are similar to each other,
and the KS test does not reveal any statistical difference between
them. The percentage of discy isophotes is larger in all regions,
except in region 4. There appears not to be a strong correlation
between \mac and ellipticity, in agreement with
\cite{2006MNRAS.370.1339H} and \cite{2014ApJ...787..102C}. The
peculiar distribution of \mac in region 4 may be the result of the
intrinsic merger history of the cluster. $N$-body simulations by
\cite{2005A&A...437...69B}, which take into account mergers of
galaxies with different mass ratios (outside a cluster
  gravitational potential), produce galaxies with boxy isophotes in
the inner part of the profile and discy ones in the outer part (i.e.,
positive radial gradients in \mac). Considering the sample studied in
this work, the profiles do not clearly show this behaviour, with half
of them showing negative gradients for the mean \mac. The Antlia
sample has a mild predominance of galaxies with \mac $> 0$ in the
innermost regions: 55\% for region 1, 54\% for region 2 (most are in
the range of $0.0 - 0.02$). This was pointed out by
\cite{2003ApJ...597..893N} as the result of binary disc galaxy
mergers, from collisionless $N$-body simulations. The larger values of
\mac may be related with hybrid mergers with very different mass
ratios \citep{2005A&A...437...69B}.

As pointed out \cite{2015MNRAS.451..791C}, the colour-magnitude
relation of the sample shows a `break' at the bright end so that
the most massive ETGs show almost constant colours. One possible
interpretation is that this is a consequence of dry mergers ---both
minor and major--- since $z \sim 2$. Then, the more massive galaxies
would evolve without gas and no further enrichment is expected
\citep{2011MNRAS.417..785J}.  The analysis of geometrical parameters
may show evidence of different possible scenarios. The largest
galaxies in the sample have regular isophotes (\mac $\sim 0$) and the
ellipticities show a wide distribution along the range: $0.0 - 0.4$,
while dEs have large deviations from perfect isophotes. Numerical
simulations of multiple mergers presented by
\cite{2007A&A...476.1179B} show that the merger remnants tend to be
boxy for 1:1 mergers \citep[see also][for dissipationless
  simulations]{2003ApJ...597..893N, 2006MNRAS.369..625N}, while larger
mass ratio (like 3:1 and 4:1) mergers result in discy-shape
ellipticals. \citet[][and references therein]{2006ApJ...636..115P}
found that discy galaxies have higher ellipticities in the sample that
they studied. On the other hand, boxy galaxies have larger half-light
radii, and tend to be bigger and brighter than discy
galaxies. \cite{2014ApJ...787..102C} and \cite{2014RAA....14..144H}
\citep[also reported by][]{2006ApJS..164..334F} found that their
sample shows a trend between $a_{4}/a$ and absolute magnitude in the
$i$-band, that could be considered similar to the boxiness trend found
by \cite{2007A&A...476.1179B} for the remnants of multiple minor
mergers, with boxiness increasing with mass ratio. In particular, we
could not confirm any relation between \mac and magnitude in our
sample. In any case, it is clear that early-type dwarfs display a
broad range in \mac at all radii, from fairly discy
to boxy shapes. This could be due to dwarfs being more strongly
affected by interactions, and/or to a mixture of objects with
different origins/histories among low-luminosity systems.

The relations between the S\'ersic index $n$ and $a_{4}/a$, $a_{3}/a$
and $\epsilon$ have been studied by different authors on different
magnitude ranges \citep{2006MNRAS.370.1339H, 2014RAA....14..144H}, who
found only a mild correlation between $n$, $\epsilon$ and
$a_{4}/a$. We found a correlation between these parameters just for
the innermost radial range; this behaviour still holds when we only
consider the faintest galaxies in the sample. We also point out that
the relatively broad ranges spanned by the values of the Fourier
parameters of dEs, cannot be explained just by the larger errors
present in the relations depicted in Figure~\ref{fig:e_n_M-a4}. Thus,
it may be an intrinsic characteristic for the fainter galaxies, which
has been shown to include several structural sub-classes pointing to
different origins \citep[and references therein]{2005MNRAS.356...41C,
  2007ApJ...660.1186L}.

\section{Summary}\label{sec:summary}%
We present the isophotal analysis as well as the surface photometry
data (catalogue) for a sample of 138 early-type galaxies in the Antlia
cluster. The scaling relations followed by them have been described in
a previous companion paper \citep{2015MNRAS.451..791C}. Our study is
based on MOSAIC\,II--CTIO images of four adjoining and slightly
superimposed fields, covering each one 36\,arcmin\,$\times$\,36\,arcmin,
and taken with the Kron-Cousins $R$ and Washington $C$ filters.

We have used ELLIPSE within IRAF to obtain the geometrical parameters
that characterize the isophotes of each galaxy along its radius. Then,
we obtained mean values of ellipticity and Fourier coefficients
$a_{3}$ and $a_{4}$ in four radial bins, weighted by the intensity of
each isophote. Total integrated magnitudes were obtained by fitting
single S\'ersic models to the observed surface brightness profiles. In
addition to presenting the surface-photometry catalogue, our main goal
was to find possible correlations among global properties. We also
looked for statistical differences between the isophotal shapes in the
inner and outer regions of the profiles, since it is supposed that
physical processes ruling the evolution of galaxies affect differently
both regions \citep[and references therein]{2014ApJ...787..102C}.
Most of the galaxies in our the sample have discy isophotes, but they
tend to change along radius, turning into boxy. The processes involved
in the evolution of the galaxies are presumably different: while in
the inner part they must be driven by internal ones, the outer regions
are more sensitive to the environment (ram-pressure stripping, galaxy
harassment, etc.) as suggested by \cite{2012ApJS..198....2K}.

\section*{Acknowledgements}
We thank to an anonymous referee for constructive remarks.  This work
was funded with grants from Consejo Nacional de Investigaciones
Cient\'ificas y T\'ecnicas de la Rep\'ublica Argentina, Agencia
Nacional de Promoci\'on Cient\'ifica y Tecnol\'ogica, and Universidad
Nacional de La Plata (Argentina). JP~Calder\'on and LPB are grateful
to the Departamento de Astronom\'ia de la Universidad de Concepci\'on
(Chile) for financial support and warm hospitality during part of this
research. MG acknowledges support from FONDECYT Regular Grant
No. 1170121. LPB and MG: Visiting astronomers, Cerro Tololo
Inter-American Observatory, National Optical Astronomy Observatories,
which are operated by the Association of Universities for Research in
Astronomy, under contract with the National Science Foundation.

\bibliographystyle{mnras}
\bibliography{complete_manuscript}
\bsp % ``This paper has been produced using the ...''
\label{lastpage}

\end{document}